\documentclass[final,1p,times,authoryear,12pt]{elsarticle}

\usepackage{amssymb}

\usepackage{longtable}
\usepackage{paralist}
\usepackage{edcomms}
\usepackage{booktabs}
\usepackage{float}
\restylefloat{table}
\usepackage{hyperref}
\usepackage{pdflscape}
\usepackage{color}
\usepackage{xspace}
\usepackage{fullpage}
\usepackage{tabularx}
\usepackage{wrapfig}

\newif\ifcomments\commentsfalse

\ifcomments
\newcommand{\authornote}[3]{\textcolor{#1}{[#3 ---#2]}}
\newcommand{\todo}[1]{\textcolor{red}{[TODO: #1]}}
\else
\newcommand{\authornote}[3]{}
\newcommand{\todo}[1]{}
\fi

\def\CC{{C\nolinebreak[4]\hspace{-.05em}\raisebox{.4ex}{\tiny\bf ++}}\xspace}

\begin{document}


\begin{frontmatter}



\title{State of the Practice for GIS Software}


\author{W.\ Spencer Smith, D.\ Adam Lazzarato and Jacques Carette}

\address{Department of Computing and Software \\
McMaster University \\
Hamilton, Ontario, Canada}

\begin{abstract}
  We present a reproducible method to analyze the state of software development
  practices in a given scientific domain and apply this method to Geographic
  Information Systems (GIS).  The analysis is based on grading a set of 30 GIS
  products using a template of 56 questions based on 13 software qualities.  The
  products range in scope and purpose from a complete desktop GIS systems, to
  stand-alone tools, to programming libraries/packages.  The final ranking of
  the products is determined using the Analytic Hierarchy Process (AHP), a
  multicriteria decision making method that focuses on relative comparisons
  between products, rather than directly measuring qualities.  The results
  reveal concerns regarding the correctness, maintainability, transparency and
  reproducibility of some GIS software.  Three recommendations are presented as
  feedback to the GIS community:
\begin{inparaenum}[i\upshape)]
\item Ensure each project has a requirements specification document;
\item Provide a wealth of support methods, such as an IRC (Internet Relay
  Chat) channel, a Stack Exchange tag for new questions, or opening the issue
  tracker for support requests, as well as the more traditional email-based
  methods; and,
\item Design product websites for maximum transparency (of the 
 development process); for open source projects, provide a developer's guide.
\end{inparaenum}
\end{abstract}


\end{frontmatter}


\section{Introduction}\label{introduction}

This paper analyzes the state of development practice in Geographic Information
Systems (GIS).  The scope and purpose of the software analyzed ranges from
complete desktop GIS systems, to stand-alone products, to programming libraries.
GIS software requires sophisticated data structures and image processing
algorithms.  The complexity of GIS software raises concerns for software
qualities such as correctness, reliability and performance.  To address these
concerns, and produce high quality software, requires solid Software Engineering
(SE) and Scientific Computing (SC) development practices.

The authors of this paper are not GIS experts; however, we are experts in SE
applied to scientific computation software.  As outsiders, we can claim
objectivity, since we have no prior attachment to any of the software examined
in this paper.  We hope to provide valuable feedback to the GIS community to
help improve the quality of their software.

We arrive at our feedback and conclusions through a reproducible process of
systematically grading software products in the GIS domain based on 13 software
qualities.  We do not grade the products on functionality.  Rather, we grade the
development process of the projects, and determine how well the projects adhere
to SE principles and practices.  A main goal of the software grading process is
objectivity, and quantification, wherever possible.  An external list of
software products written by a domain expert acts as an authoritative list of
software to be graded.  As a part of the grading, we preform a pairwise
comparison between each of the software products using a multicriteria decision
analysis process.  Based on the rankings from the decision analysis, we then
analyze trends between software products.

Our inspiration for this project comes from \citet{gewaltig-neuroscience}, and
the later paper, \citet{gewaltig-neuroscience2}. (Our work is based mainly on
the earlier version, since the classification system in that paper is the
simpler of the two, and it still fulfills our needs).  In their papers, Gewaltig
and Cannon perform a software review in the domain of computational
neuroscience.  We build and expand on their process, as we previously did for
mesh generation software~\citep{SmithEtAl2016} and for seismology
software~\citep{SmithEtAl2017SS}.  Gewaltig and Cannon's review gathers data
from the public information on the software product's website and analyzes it
for trends to build feedback.  The authors conclude that there is often a
misunderstanding between developers and users regarding the reasons for creating
the software and expectations for its features.  Much of the software examined
was written by students during their Master's or PhD research; many of the
developers did not have backgrounds in computer science or SE.  Their priority
was their scientific application, not best practices in SC. \citep{Segal2007}
refers to this category of developers as professional end user developers.  This
type of developer seems common for SC software.

One major problem with scientific software development is a communication
barrier between scientists and software engineers when discussing requirements.
The barrier exists because the scientists have experience in their field, but
less with software development.  The scientists know that the requirements will
change, but they cannot precisely convey how they will evolve. Not correctly
articulating requirements, or changing requirements midway through a project,
greatly impacts the productivity of the development team \citep{segal-case}.
When engineers create the software, the resulting development artifacts, such as
user manuals and introductory examples, are not sufficient for the scientists to
understand the product \citep{segal-developing}.  When end users (scientists)
develop the software product, the situation is not improved, since their
training in science has not prepared them to consider important software
qualities, like maintainability and reusability.  The differences between SE and
SC has led to a chasm between these two disciples \citep{Kelly2007}.

The remainder of this article is organized as follows: Section~\ref{background}
provides background information and outlines previous work.  Our methods are
explained in Section~\ref{methods}.  A summary of our results is presented in
Section~\ref{results} and our recommendations are detailed in
Section~\ref{recommendations}.  Concluding thoughts are found in
Section~\ref{conclusions}.

\section{Background} \label{background}

The definitions found in this section include the software qualities and SC best
practices that our software grading template is based on. Also included is a
quick overview of the Analytic Hierarchy Process (AHP), a multicriteria decision
making method that we use to analyze the results of the software grading.

\subsection{Software Qualities} \label{Sec_SoftwareQualities}

Our analysis is built around a set of software qualities.  Software qualities
can be \textit{internal}, in which case the qualities only concern developers,
or \textit{external}, in which case the qualities are visible to the end users
\citep[p.~16]{ghezzi-se}.  Strong internal software qualities help achieve
strong external qualities.  Qualities not only concern the software
\textit{product} itself, but also the \textit{process} used, and the artifacts
generated \citep[p.~16--17]{ghezzi-se}.  Artifacts include documentation and
test data, which are created to improve and measure the software's quality.

This paper measures 13 software qualities, as summarized
in~\citet{SmithEtAl2016}: installability, correctness and verifiability
(measured together), reliability, robustness, performance, usability,
maintainability, reusability, portability, understandability, interoperability,
visibility and reproducibility.  The majority of the above \textit{qualities of
  software} come from\citet{ghezzi-se}.  We have excluded qualities that we
would not be able to sufficiently measure, such as \textit{productivity} and
\textit{timeliness}.  We have also added two qualities that we believe are
important to the overall quality of SC software: \textit{installability} and
\textit{reproducibility}.

The above software qualities come from SE, and apply to any class of software.
Specific SC software development principles are also important to consider when
examining GIS software.  The ``best practices'' \citep{wilson-best-practices}
for SC form a checklist of eight basic practices that promote reliable and
maintainable code. We use the key ideas from this checklist for creating our
grading template. For example, from this list we draw our standards for source
code documentation, reuse of libraries, the use of a issue tracker and other key
elements of correctness and maintainability.

\subsection{Analytic Hierarchy Process}

The Analytic Hierarchy Process (AHP) is a multicriteria decision making process.
The objective of AHP is to compare multiple results based on multiple criteria
important to the decision \citep{saaty-ahp}.  In this paper, AHP is used in part
to compare qualities of software between each other. Since there is no formal
scale or units in which these qualities are measured, we use AHP to remove this
problem and focus on relative and pairwise comparisons.  AHP consists of a set
of $n$ \textit{options} and a set of $m$ \textit{criteria} with which the
options are graded. The criteria are prioritized.  Then, for each of the
criterion, a pairwise analysis is performed on each of the options, in the form
of an $n$x$n$ matrix $a$.  The value of $a_{jk}$ ranges from 1, when options $j$
and $k$ are equally graded, to $9$, when option $j$ grades extremely (maximally)
higher than $k$.  The definitions of the values between 1 and 9 are found in
\cite{saaty-ahp}.

The value of $a_{kj}$ is the inverse of $a_{jk}$ ($a_{kj}=1/a_{jk}$). Once the
matrix $a$ has been filled, weights are generated by creating a new matrix $b$.
Entry $b_{jk} = a_{jk}/\sum(a_{\cdot k})$, where the dot ($\cdot$) indicates the
entire row. Next, these weights are averaged to determine the overall score for
that option and criterion. All of the scores are weighted according to the
priorities of the criteria. Final scores are generated to create a ranking for
each of the options. These final scores give a high-level view of how an option
compares to the others based on all criteria \citep{mocenni}.

In our project, the $n$ graded software products are the options. The 13
software qualities are the $m$ criteria. In \citet{trianta-ahp} the authors warn
that options' final scores should not be considered as absolute ranks. In our
experiment, we certainly do not wish to absolutely rank software products, but
more to sort the software products into groups based on their software
qualities.

\section{Methods}\label{methods}

In this paper, we create a systematic grading and analysis procedure for a list
of SC software products, in particular GIS software. First, the software is
graded, based on the software qualities and best practices of SC software from
Section~\ref{background}.  Second, the results are discussed and analyzed for
trends.

\subsection{Software Product Selection}

To select the software for analysis, we followed John W.\ Wilson's list of
``Useful remote sensing software'' \citep{birder}.  The list provides a
comprehensive list of GIS software, and libraries.  Most of the software is free
and open source, with contributions from both researchers and independent
developers.  Not all of the links to software products in Wilson's list were
used.  For example, links to the Python programming language and the R project
for statistical computing, were removed because these are general programming
languages, and thus not specific to GIS.  Additionally, the links to sample data
sets and tutorial web pages were not considered, since they are not software
products.  The full list of software graded can be found in
Section~\ref{results}.  In total, there are 30 software products on the list.

\subsection{Grading Template}

The template we used for grading the software products is a collection of 56
questions.  The full list is available in the Appendix and at
\url{https://data.mendeley.com/datasets/6kprpvv7r7/1}.  The questions are divided into the
13 software qualities listed in Section~\ref{Sec_SoftwareQualities}.  Due to the
qualitative or subjective nature of some of the software qualities
(e.g.~reliability, robustness), the template had to be carefully structured.
When choosing questions (measures), we aimed for unambiguity, and quantification
where ever possible (e.g.~yes/no answers).  As outsiders, we looked for measures
that are visible, measurable and feasible in a short time with limited domain
knowledge.  Unlike a comprehensive software review, this template does not grade
on functionality and features.  Therefore, it is possible that a relatively
featureless product can outscore a feature-rich product.

In the first section of the template, general information is gathered about the
software. This information contains the software name, URL, license information,
possible educational backing, funding methods, and the dates for when the
project was released and when it was last updated. A project is defined as
\textit{alive} if it has been updated within the last 18 months, and
\textit{dead} otherwise. This time frame is arbitrary, but it seems appropriate
since this includes the usual time frame for new operating system updates and
more than a full calendar year for educational institutions. As per
\citet{gewaltig-neuroscience}, we define the category of \textit{public}
software as software intended for use by the public. \textit{Private} (or
\textit{group}) software is only aimed at a specific group of people. Lastly,
\textit{concept} software is available simply to demonstrate algorithms or
concepts, and not for use in a production setting. The main categories of
development models are: \textit{open source}, where source code is freely
available under an open source license; \textit{freeware}, where a binary or
executable is provided for free; and, \textit{commercial}, where the user must
pay for the software product.  If the product is open source, we note the
programming language used.

We use a virtual machine to provide an optimal testing environments for each
software product.  During the process of grading the 30 software products, it is
much easier to create a new virtual machine to test the software on, rather than
using the host operating system and file system.  Adding and removing software
from one's computer can often be difficult; we use virtual machines to avoid
this headache. Once grading of a software is complete, the virtual machine with
the software on it is destroyed, and the host operating system is
oblivious. Virtual machines also provide fresh installs of operating systems,
which minimizes or completely removes ``works-on-my-computer'' errors.  Unless
the software has dependencies that must be installed, any installation
instructions that are provided by the software developers should be compatible
with a fresh install of an operating system. In our grading data, we note the
details of the virtual machine, including hypervisor and operating system
versions.

\subsection{Measuring Qualities}

For each of the following qualities, the software receives a grade from one to
ten.  These grades are the grader's subjective feeling about the software based
on the measurements, past experiences and the other GIS software products.  The
grader must aim for consistency during the grading process.  At the end of the
ranking process, the potential subjectivity is mitigated by the use of AHP,
since in AHP it is the relative difference that matters.  As long as two graders
are internally consistent, with their grades mostly trending in the same
direction, their relative comparisons matrix in AHP should be similar.  The
objectivity of the grading process is discussed further in
Section~\ref{SecApproachToGrading}.

\textit{Installability} is an aspect of the software that we can thoroughly
analyze.  To grade qualities such as usability or robustness, we must first
install the software.  Installation is also the primary entry point for every
user of the software: beginner or advanced. We check for the absence or presence
of install instructions.  These instructions are ideally linear and highly
automated, including the installation of any external libraries that need to be
installed.  Before proceeding, if there is a way to validate the installation,
we run the validation. At the end of the installation and testing, and if an
uninstallation is available, we run it to see if any problems were caused. The
complete grading template for installability is presented in
Table~\ref{table:template}.  A similar set of measures is used for the other
quality gradings.

\begin{table}[ht]
\begin{tabularx}{\textwidth}{p{13cm} X}
\toprule
\textbf{Installability Measure}  & \textbf{Metric}\\
\midrule
Are there installation instructions? & yes, no\\
Are the installation instructions linear? & yes, no\\
Is there something in place to automate the installation? & yes$^*$, no\\
Is there a specified way to validate the installation, such as a test suite? & yes$^*$, no\\
How many steps were involved in the installation? & number\\
How many software packages need to be installed? & number\\
Run uninstall, if available. Were any obvious problems caused? & unavail, yes$^*$, no\\
\bottomrule
\end{tabularx}
\caption{Installability grading template (unavail means that uninstall is not
  available and a $^*$ indicates that the measurement should also be accompanied
by explanatory text.)} \label{table:template}
\end{table}

\textit{Correctness} is difficult to grade because it is an absolute quality.
What we are actually measuring is confidence in correctness and the related
quality of \textit{verifiability}.  To accurately grade
correctness/verifiability, there must be a requirements specification document
and the behaviour of the software must strictly adhere to it. We do not have the
time or the means to rigorously test every piece of software to this extent. We
look for indirect means of judging correctness. For instance, we look for the
use of standard libraries, in which the community has confidence, and confidence
building techniques, such as such as assertions in the code, and documentation
generated from the code.

Since the duration of our usage of the software is quick and structured, we can
easily analyze surface \textit{reliability}. We cannot grade long term
reliability of the product, but poor reliability during grading is certainly a
cause for concern. We know how the software is expected to behave via the
installation guide and tutorial (if present), and we also complete a getting started
tutorial, if available. If the software does not behave as expected during this
duration of usage, the software is graded poorly with respect to reliability.

When we grade surface \textit{robustness}, we are trying to break the software.
We cannot test all features of a product, and we cannot provide exhaustive cases
of garbage input to the software. Purposely making errors during the getting
started tutorial and other interactions with the software tests the robustness
of the program, and its error handling.

\textit{Performance} is a very difficult quality of software to measure.  For
practical reasons, the size of the problems we are testing cannot strain the
products. Instead of measuring performance directly we look on the surface for
evidence that performance was considered, such as a makefile that shows the
presence of a performance profiler.

Surface \textit{usability} is based on our impressions of the ``human-
friendliness'' of the product during the grading time frame. During our time
using the product, we checked for the existence of a getting started tutorial.
This tutorial is an explicit guide for first time users that has linear steps
for the absolute basic usage of the product. We also look for a more detailed
user manual. If any features are hidden or difficult to find, then they do not
satisfy Norman's design principle of visibility \citep{Norman02}. We also
measure whether the software has the expected ``look-and-feel'' for products for
that platform. User support techniques, such as web forums, are also considered
when assessing usability.

\textit{Maintainability} is one of the more concrete software qualities to
grade.  Whether or not the developers write a changelog, use an issue tracking
tool, or have multiple versions of the software, are all easy things to
examine. If the developer gives information on how their code is reviewed, or
have specific instructions on how to contribute to the project, this information
adds to the maintainability of a product.

\textit{Reusability} is a strong theme in both SC best practices and in SE in
general. In our grading, we note products that are currently being reused or
that make reusability simple. Adding plugin or add-on functionality greatly
improves reusability, especially when well-documented. Also, in the case of an
API (Application Programming Interface), having full, concise documentation
available for programmers improves reusability.

\textit{Portability} is graded based on what platforms the software is
advertised to work on, and how the developers handle portability. Using cross-
platform code or a cross-platform build system is evidence that portability was
considered in the design and development. Any related discussion of portability
or build practices is also noted.

For grading \textit{understandability}, we examined the source code that comes
with open source software products.  We checked the source code for objective
properties, like modularity, consistent indentation and if concise commenting is
used.  If there exists a coding standard enforced by the project, it helps
understandability.  We also checked for documentation regarding software design,
such as a module guide for the system architecture.  If source code is
unavailable, the software is not graded on this criterion.

\textit{Interoperability} grading consists of examining if the product can
communicate or otherwise interact with any external systems. We checked for this
kind of interaction and whether an external API document is provided.

\textit{Transparency} is a quality that is ever-present when grading software
products. All information we need for grading, including the getting started
tutorial and the source code itself all depend on how the information is
presented, and how easy it is to find. We are also interested in whether a
development process has been defined. For example, a waterfall or a spiral
development model could be used, or perhaps a more ad-hoc process has been
documented.

\textit{Reproducibility} measures any evidence or documentation of development
or testing platforms. If there are any tools that alleviate inconsistencies
between hardware or software platforms, the reproducibility of the software's
results can be tested. As stated above, there are several reasons we use virtual
machines for using software during grading. These reasons are applicable for
development as well. Documented methods of development or testing on virtual
machines greatly helps reproducibility.

\subsection{Approach to Grading} \label{SecApproachToGrading}

During the grading process, the grader is faced with the task of getting a
concise snapshot of each software product, based on one to three hours of
interaction. The grader needs a good strategy to approach this task.  Each
grader may have a strategy that is slightly different from the others.  Even
though we aim for quantification wherever possible, it is unrealistic to expect
exactly the same results between all potential graders. The key is to aim for
relative consistency between graders, which should be possible since AHP is
performing pair-wise comparisons between the grades.

The process of pair-wise comparison is automated using a software program
that converts the grades (from 1 to 10 on each quality for each product) to an
AHP comparison matrix.  Once the AHP calculations are complete, we can see how
the software products grade relative to one another.  Further details on the
algorithm to transform our objectives measures into an AHP sorted ranking can be
found in \cite{SmithEtAl2015-SS-TR}, which presents an analysis similar to the
current one, except rather than studying GIS software, the domain of interest is
psychometrics software.

For grading GIS software, we award a grade of 5 for ``indifference.''  For
example, if the developer has not explicitly written about or documented extra
measures to increase the performance of an otherwise sound product. We cannot
dock marks for poor performance, and we cannot award marks for outstanding
performance. This same situation appears often in portability and reusability
grading.

We award marks of 1 for understandability (of the code) when no source code is
available, since we cannot analyze the product's understandability, so relative
to any open source product the product's understandability is very poor.

To demonstrate that the grading process is objective, 5 software products were
graded by two reviewers.  The final results were very similar, and the final
grades nearly exactly the same.  The main source of difference between gradings
is the interpretation of the definition of correctness, and specifically what a
requirements specification document entails. As long as each grader is
consistent, the relative comparisons in the AHP results will be consistent
between graders. Changes in perceived visibility of the software product also
plays a major part in differences between grades. Information that is hard to
find, or on a different site, can hurt a product's grades, since not all
reviewers will have the same luck in finding the information.

\begin{table}
\centering
\begin{center}
\begin{tabular}{p{8cm}p{1cm}p{1cm}p{2cm}}
\toprule
Name                        & Status    & Open source   & Language\\
\midrule
DIVA-GIS \citep{DIVAGIS}    & Dead      & No            & Java\\
GRASS \citep{GRASS}         & Alive     & Yes           & C\\
gvSIG \citep{gvSIG}         & Alive     & Yes           & Java\\
QGIS \citep{QGIS}           & Alive     & Yes           & \CC, Python\\
SAGA-GIS \citep{SAGAGIS}    & Alive     & Yes           & \CC\\
uDig \citep{uDig}           & Alive     & Yes           & Java\\
\bottomrule
\end{tabular}
\end{center}
\caption{Desktop GIS set} \label{table:gis}
\end{table}

\section{Summary of Results}\label{results}

The most up-to-date and complete grading of the 30 domain software products is
available in an external repository at
\url{https://data.mendeley.com/datasets/6kprpvv7r7/1} with a less verbose
summary available in the Appendix.

Before grading the software qualities, we gathered general information about the
products.  Of the 30 GIS products, eight were associated with educational
institutions. These institutions are the workplaces of the developers, or
provide support for the projects financially. There are 19 open source
products. The 30 software products are easily split into three main sets. These
sets will be used to simplify the presentation of the software products
throughout the remainder of this paper. First, there are six \textit{Desktop
  Geographical Information System (GIS)} products, as shown in
Table~\ref{table:gis}. These products have enormous feature sets and exist to
obtain, change, analyze and present a wide variety of geographical data. The
next set consists of 12 \textit{stand-alone tools} (Table \ref{table:so-tools})
that perform specific tasks. These tools are much less feature-rich than the
desktop GIS systems. Finally, there are 12 \textit{libraries} and
\textit{packages} (Table~\ref{table:prog-lib}) that enable programmers to
develop their own software products using the functionality of the
libraries/packages. Of these libraries, seven are written in Python, three in R,
one in C, and one in \CC.

\begin{table}[p]
\centering
\begin{center}
\begin{tabular}{p{7.5cm}p{1.5cm}p{1cm}p{2cm}}
\toprule
Name                                & Status    & Open source   & Language\\
\midrule
Biomapper \citep{Biomapper}         & Dead      & No            & Borland Delphi\\
Conefor \citep{Conefor}             & Dead     & Yes           & \CC\\
CROP\_VGT \citep{CROPVGT}           & Dead      & No            & Unclear\\
CyberTracker \citep{CyberTracker}   & Alive   & No            & \CC, Java\\
DesktopGarp \citep{DesktopGarp}     & Unclear   & No            & \CC\\
FRAGSTATS \citep{FRAGSTATS}         & Dead     & No            & \CC\\
Lifemapper \citep{Lifemapper}       & Unclear     & Yes           & Python\\
MARXAN \citep{MARXAN}               & Dead      & No            & \CC\\
Maxent \citep{Maxent}               & Alive      & No            & Java\\
openModeller \citep{openModeller}   & Dead     & No            & \CC\\
OSSIM \citep{OSSIM}                 & Alive     & Yes           & \CC\\
Zonation \citep{Zonation}           & Dead     & No            & \CC\\
\bottomrule
\end{tabular}
\end{center}
\caption{Stand-alone tools set} \label{table:so-tools}
\end{table}

\begin{table}[p]
\centering
\begin{center}
\begin{tabular}{p{8cm}p{1cm}p{1cm}p{2cm}}
\toprule
Name                        & Status    & Open source  & Language\\
\midrule
GDAL/OGR \citep{GDALOGR}    & Alive     & Yes           & \CC\\
GDL \citep{GDL}             & Alive     & Yes           & \CC\\
geopy \citep{geopy}         & Alive     & Yes           & Python\\
landsat \citep{landsat}     & Dead     & Yes           & R\\
NetworkX \citep{NetworkX}   & Alive     & Yes           & Python\\
NumPy \citep{NumPy}         & Alive     & Yes           & C, Python\\
PostGIS \citep{PostGIS}     & Alive     & Yes           & C\\
pyproj \citep{pyproj}       & Alive      & Yes           & Cython\\
pyshp \citep{pyshp}         & Alive      & Yes           & Python\\
raster \citep{raster}       & Alive     & No            & C\\
rgdal \citep{rgdal}         & Alive     & Yes           & C, \CC, R\\
shapely \citep{shapely}     & Alive     & Yes           & Python\\
\bottomrule
\end{tabular}
\end{center}
\caption{Programming libraries set} \label{table:prog-lib}
\end{table}

Summary general information about the graded GIS software follows:

\begin{itemize}
\item 17 products have 5 or fewer developers. Eight projects have two or fewer
  developers.
\item 5 products (GRASS, gvSIG, QGIS, OSSIM, PostGIS) have funding by The Open
  Source Geospatial Foundation \citep{OSGeo}, whose goal is to support the
  development of open source geospatial software products.
\item There are 9 dead products based on our 18 month time-frame for liveness.
\item Of the 19 open source products, the GNU GPL license is the most popular
  (11/19). MIT (4/19) and BSD (3/19) licenses are also widely used. Closed
  source software, or ``freeware'' either provide no license, and explicit
  written terms of use, or an end user license agreement.
\item Windows is well supported (29/30).
\item \CC is the most popular language, in use by 13/30 products.
\end{itemize}

\begin{wrapfigure}{R}{0.6\textwidth}
\includegraphics[width=0.59\textwidth]{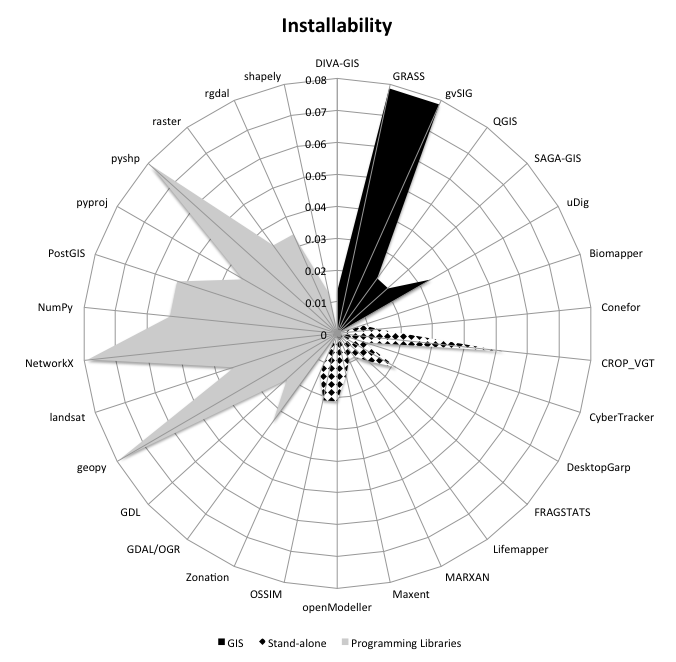}
\caption{AHP results for installability.}
\label{fig:ahp-installability}
\end{wrapfigure}
With respect to \textit{installability}, 21/30 projects contained installation
instructions, with 14 of the 21 having linear instructions. Therefore, more than
half of the products analyzed did not contain linear installation instructions.
Though, in 27/30 cases, installation was automated with the use of makefiles or
scripts. So, the absence of linear installation instructions is partially
justified in that the steps are taken automatically.  Only two software products
(NumPy and PostGIS) provided explicit post-installation tests to check the
correctness of the installation. Eight products required software to be
installed beforehand.  Uninstallation automation is not provided in 13 of the 30
projects. However, deleting the software's root directory or deleting the
executables was normally sufficient to uninstall the software.

As Figure~\ref{fig:ahp-installability} shows, programming libraries generally do
well on installability.  This is because these products can often be installed
using just one step (9/12) using a package manager, like pip for Python software
and CRAN for R software. GRASS and gvSIG, from the Desktop GIS set, also score
high on installability, since these products include easy to use installers and
have easy-to-follow, linear installation instructions.  Poor cases of
installability occur when the user must ``jump hurdles'' to obtain or install
the software. Problems occur when users must do additional research, or follow
an installation practice that requires an extra layer of software or
``work-arounds'' . For example the only supported way to install DIVA-GIS
(native on Windows) on OS X is through Winebottler, an .exe packager for OS
X. This is not a ``normal'' way to support OS X and relies on a third party for
installation and portability. This is different from using a virtual machine for
installing and using the software, as we have done for our measurement
purposes. In the case of MARXAN and Conefor's command line tools, personal
information (name, email) or email correspondence with the developers is
required to obtain the software. Developers have every right to ask for this
information before making the software available to the user, but this still
adds complexity to the installation process.

\begin{wrapfigure}{R}{0.6\textwidth}
\includegraphics[width=0.59\textwidth]{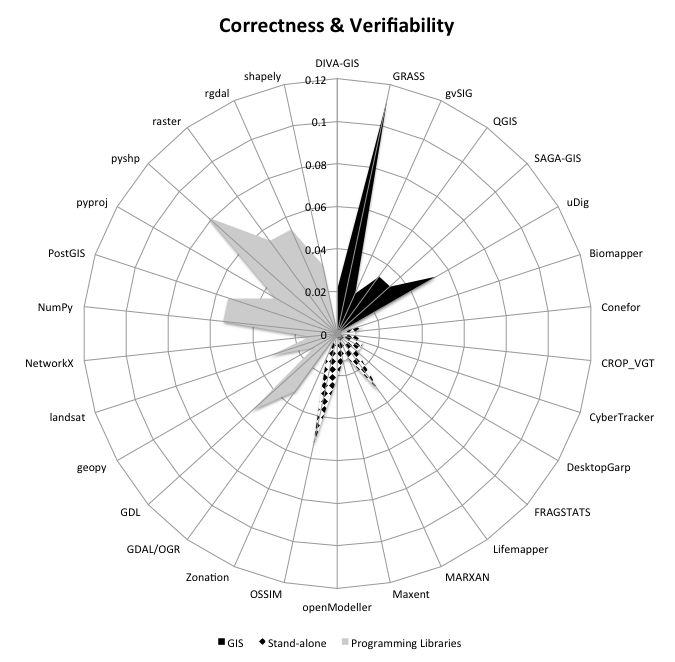}
\caption{AHP results for correctness \& verifiability.}
\label{fig:ahp-correctness}
\end{wrapfigure}

As Figure~\ref{fig:ahp-correctness} shows, \textit{correctness and
  verifiability} score high for programming libraries and some desktop GIS
systems, but not particularly well for the stand-alone tools. 18 systems used
external libraries, with the stand-alone tools using external libraries less
frequently than the desktop GIS or programming library sets. Some of the most
relied-upon software include sp (written in R), GDAL and PostGIS for abstracting
the handling and storage of spatial data. Requirements specification documents
were very rare. Only three products (GRASS, GDL and pyshp) explicitly stated
adherence to a specification. In GRASS, this specification is presented in a
wiki that outlines the purpose, scope, overall description and specific
requirements, such as performance and design constraints. While specification
documents often do not exist, some projects contain other evidence of explicitly
considering correctness. Doxygen or similar tools are used in projects such as
SAGA-GIS, PostGIS and OSSIM to automatically generate documentation from the
source code. This adds to correctness because the specified behaviour of the
product is derived from the source code, and by maintaining them together the
documentation and the code should be in sync.  Another form of confidence
building is automated testing. Five desktop GIS systems, and 10 programming
libraries used automated testing. Though stand-alone tools show a general lack
in automated testing (2/12).  Without testing, requirements specification or
other evidence, the conclusion is that stand-alone tools have not adequately
considered the quality of correctness.

\textit{Reliability}, overall, is very strong on the surface for all three sets
of software products. As explained above, installation of the 30 products went
smoothly for the most part. Terminal errors or other prohibitive problems during
installation were rare and only occurred in two products: Lifemapper and OSSIM
(both stand-alone tools). Initial testing of the products was less automated,
and contained more room for error, especially if there are multiple steps in the
getting started tutorial. There were no errors or other ``breakages'' for
desktop GIS products or stand-alone tools during initial testing.  However, a
programming library, geopy, had a segfault error while running the getting
started tutorial.

Surface \textit{robustness} is considered in all 30 of the software products
graded. By making simple typos and using purposely broken/poor input, we were
successful in triggering errors in the software products, without the product
crashing. All of the software contained some form of error handling, with
variations on the style of display and amount of information given in the error
message. These variations impact the usability of the product. Good information
and prominent placement of blocking errors helps the user understand the
errors. While the software all contains error handling, some software products,
like raster, give difficult to understand and vague error messages, which
provided little information on what the error was or how to proceed. On average,
programming libraries performed better than desktop GIS systems and stand-alone
tools, giving more informative and noticeable errors on the command line, as
compared to the various methods of displaying errors in a GUI environment.

Surface \textit{performance} is not explicitly discussed in 22/30 software
products. GRASS and QGIS have sections of documentation related to performance.
This documentation covers performance optimization measures taken by the
developers, and/or benchmarks using test data. Performance is considered in one
stand-alone tool (openModeller) in a document detailing the methods to profile
the product's performance. In the case of programming libraries, the task of
achieving maximum performance lies with the end user. Just one programming
library (PostGIS) had any documentation about performance considerations.
Sometimes, like in the case of FRAGSTATS and GDAL/OGR, the only time performance
is mentioned is when there are possible known memory leaks or other performance-
related bugs.

\begin{wrapfigure}{L}{0.6\textwidth}
\includegraphics[width=0.59\textwidth]{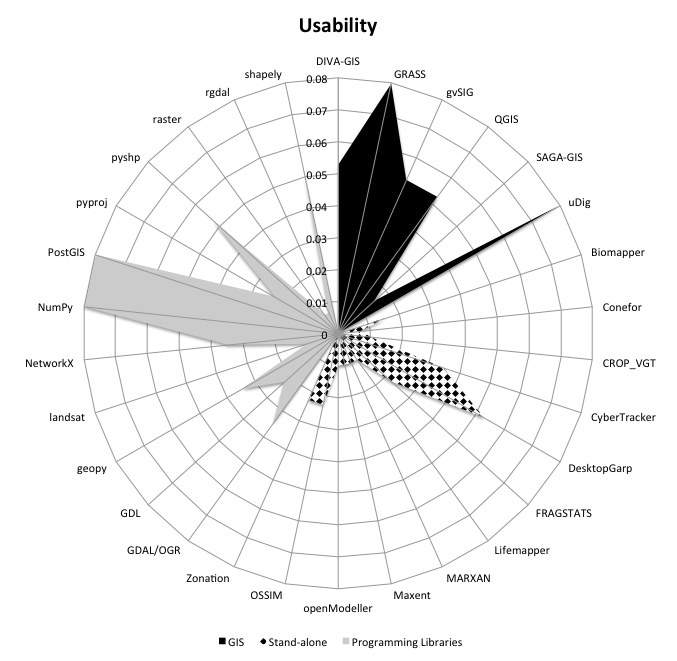}
\caption{AHP results for usability.}
\label{fig:ahp-usability}
\end{wrapfigure}

Surface \textit{usability} is strong for both desktop GIS products and
programming libraries, as shown in Figure~\ref{fig:ahp-usability}. Desktop GIS
products contained a getting started tutorial most frequently, with 4/6
products, compared to 5/12 for stand-alone tools and 7/12 for programming
libraries. These getting started tutorials normally contain a standard example
directed toward first-time users. 29/30 of the software products contained a
complete user manual. These user manuals vary in scope and length, but serve to
inform the user of the software's complete purpose, design and
functionality. The best user manuals come with the desktop GIS products. Their
user manuals are logically organized into sections that cover all of the user's
interactions with the product, from pre-installation information (feature
overview, marketing) to software design, to information and guides on using
every facet of the software.  The one product that does not contain a user
manual is CROP\_VGT. In this case, the getting started tutorial serves as the
complete guide on how to use the software. Some of the user guides are more
academic, such as the documentation for MARXAN, which consists of references to
books, and external manuals written by others.

The layout and design of the software products is encompassed by usability.
Design is more apparent in GUI applications, but design considerations are also
apparent in command line software or programming libraries. For the most part,
the expected ``look and feel'' of the software products are adhered to. Rarely,
(eg. Biomapper, OSSIM) some unusual choices are made for such things as the
font, or the GUI skin.  There are a few rare cases where the design of the
software makes important features more difficult to find than they could be.
For instance, there is a lack of organization on the settings screen in Maxent.
This is known as a problem with visibility, as described in Don Norman's design
principles \citep{Norman02}.

In most cases, (27/30) the expected user characteristics are not documented.
Conefor advises that you should be an advanced user to use the command line
tools, but otherwise, developers rarely document any background knowledge that
potential users should have before using the product.

Another aspect of usability to consider, is the existence of a user support
system. Other than direct email, and the issue tracker (if it exists, and if it
is used for posting support requests) there are alternate methods of support,
such as mailing lists, IRC (Internet Relay Chat), message boards, and FAQs
(Frequently Asked Questions).  Most frequently, 4/6 desktop GIS products had
alternatives, like an IRC channel for uDig and a dedicated QGIS StackExchange
tag for questions. Stand-alone tools (7/12) and programming libraries (6/12)
also had alternatives for support. Deviating from the norm, some projects
written in R, like raster, had no explicit support model.

\begin{wrapfigure}{R}{0.6\textwidth}
\includegraphics[width=0.59\textwidth]{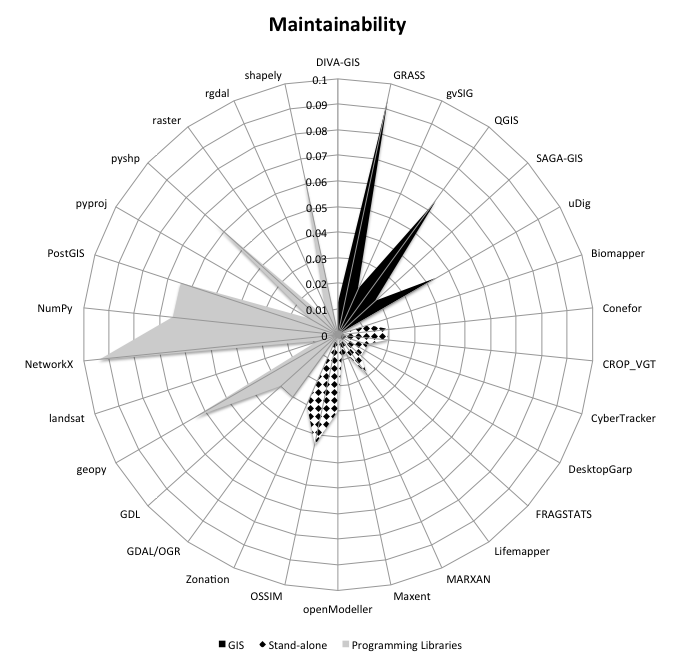}
\caption{AHP results for maintainability.}
\label{fig:ahp-maintainability}
\end{wrapfigure}

\textit{Maintainability} roughly varies with the size of the project.  With the
information available, size is difficult to quantify; however, the grader can
form a feel for the size of the project from the number of developers and
downloads, and the activity in the news sections and support channels. Based on
the reviewer's feel for project size, larger projects generally perform better
on maintainability. The developers of small or closed-source projects do not
always consider maintainability, as shown in
Figure~\ref{fig:ahp-maintainability}. 29 products had multiple versions of the
software, but often these past versions were not available for download. The
user may not ever want to download these legacy versions, but having them
available does not hurt, and improves visibility. 14 of the software products
did not use an issue tracking tool, or asked for email correspondence to report
bugs. Email correspondence is private, so the reported bugs are not known to
all, which is bad for both visibility and maintainability. Of these 14 products,
10 are stand-alone tools. Out of the remaining 16 products that are using issue
tracking, 14 of them were mostly dealing with corrective maintenance. Desktop
GIS systems (5/6) and programming libraries (9/12) mostly used issue tracking
tools.  Stand-alone tools only used issue tracking in 2/12 cases (OSSIM and
openModeller). When issue trackers were employed, the majority of the tickets
have been closed for most products. Trac, GitHub, JIRA, and Sourceforge are the
most popular issue tracking systems.

Version control systems are publicly used in desktop GIS (5/6) and programming
libraries (9/12), but again, in just 2/12 cases for stand alone tools (OSSIM and
openModeller). The developers of any of the graded products may be privately
using version control systems, but there is no documentation suggesting so. Git
(10) and SVN (7) are nearly equal in use among the graded software products.

The best cases of maintainability come from software products with
\textit{developer's guides}. Four desktop GIS systems and two programming
libraries contained developer's guides. Any information associated with the
process of adding new code to the project from internal or external contributors
can be included in a developer's guide. The software products that contain
developer's guides are: GRASS, gvSig, NetworkX, PostGIS, SAGA-GIS, uDig.

As an alternative to explicitly documenting the development process, the process
can be implicit in the workflow of the tools employed.  For example, products
that use GitHub adhere to the processes of the Git version control system and
the pull request system facilitated by GitHub.

\textit{Reusability} scored high for desktop GIS and programming library
products, but for different reasons. Five out of six desktop GIS systems contain
ways to make reusability easy using APIs, and add-ons.  An outstanding example,
GRASS GIS, contains an API and an add-ons system. These systems provide the
software product's functionality to developers so that they can create their own
functionality both inside GRASS and in their own programming projects.
Programming libraries, on the other hand, provide reusability because the
software product itself is the code and available for programmers to use for
their own purposes.

For both desktop GIS and programming library software, documentation is
important.  Well-written and designed add-on API documentation can makes it
easier for developers to learn how to interact with the products. For stand
alone tools, reusability does not seem to be a primary concern. For these
products, the developers either do not have the resources or requirements to
develop an API or plug-in system.

\textit{Portability} has been achieved for most of the software products, with
29 products supporting Windows and of these, 22 supported Linux, OS X or both.
There are 7 Windows-only products. There exist many different ways to achieve
portability including cross-platform build systems such as cmake, OS-specific
branches in code, or by the use of a language easily compiled or interpreted on
different systems. For example, languages like R and Python can be run on any
modern OS. Therefore, the programming libraries set is graded very well on
portability. In some cases, portability was explicitly identified as not being
important, which means a lack of portability cannot be held against these
products, since they have matched their own stated requirements. SAGA-GIS stated
that support for OS X is possible, but the developers had not tested
it. DesktopGarp explicitly stated that there are no plans for OS X/Linux
support.

\textit{Understandability} of the code, overall, is strong on the surface for
all sets of software products. We examined the 19 open source products' source
code and found consistency in formatting, and in the cases of products with
developers manuals, sometimes even code style guidelines (uDIG, OSSIM, NumPy) or
formatting tools (PostGIS). Useful commenting is almost always used. In one case
(pyproj), the source contains little formatting, and the grades were lowered
accordingly.  For larger projects such as the desktop GIS products, and
particularly ones with developer's guide, there are often design documents. Nine
open source projects had a design document as a reference.

\textit{Interoperability} is similar to reusability in that projects that
require these facilities often support them well. This occurs primarily in the
desktop GIS and programming library sets. These sets use external libraries more
frequently, and support re-use via add-ons or directly via an API. For example,
geopy is an API itself, but geopy interacts with many external services such as
Google and Mapquest to obtain geocoding data.

\begin{wrapfigure}{R}{0.6\textwidth}
\includegraphics[width=0.59\textwidth]{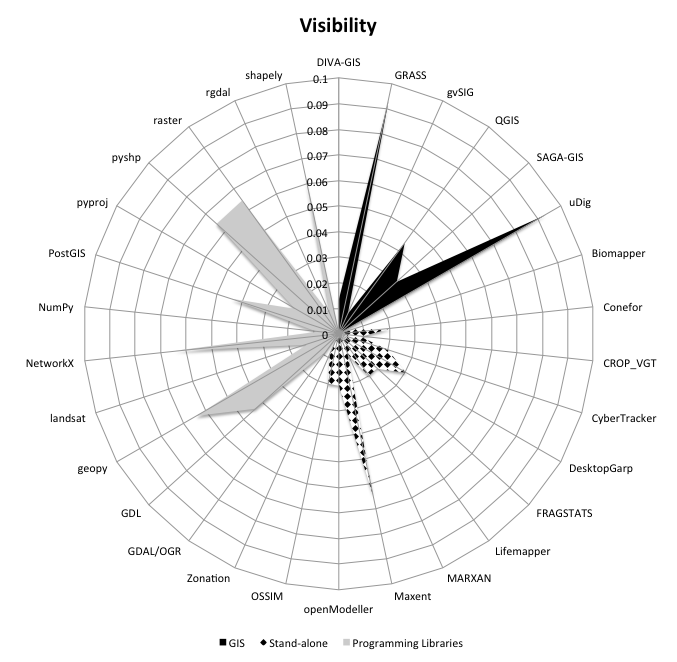}
\caption{AHP results for transparency.}
\label{fig:ahp-visibility}
\end{wrapfigure}

\textit{Transparency} seems to be roughly proportional to the size of the
project, as illustrated in Figure~\ref{fig:ahp-visibility}. The more information
the project has to display, the more often the developers have designed
efficient ways to access this information. Ideally, projects have one web site
with all information contained on it. In practice, projects often consist of
multiple web sites that provide different services for the project. For example,
a main site serves as a hub to external code hosting, download sites, issue
trackers, and/or documentation. In this case, it is the grader's task to
discover these web sites and gather information about the project.

Key to the transparency of a product is whether its development process is
defined. Any protocols that the developers use to add new code, keep track of
issues or release new versions are ideally recorded, so that new developers, or
users, can be informed. Ten projects had defined development processes. The most
thorough information regarding development process was found in the developer's
guides. These guides cover development processes, software design, code style
and more. Only 8 projects contained any developer-specific documentation section
with 7 of them having explicit developer's manuals. Of these 7, 5 were desktop
GIS products and the other 2 were programming libraries. Six of the 7 projects
with developer's manuals have 5 or more developers. The desktop GIS set has
excellent transparency, since these projects have large groups of developers to
coordinate.  Stand alone tools normally use self-made sites, so the relative
transparency can vary, but, in general, this set of GIS software graded poorly
in transparency, especially tools like Biomapper or Zonation.

Open source programming libraries can rely on code hosting services such as
GitHub or SourceForge to consolidate information and tasks such as issue
tracking and a wiki. Software packages available via repositories, such as R
software in The Comprehensive R Archive Network (CRAN), can be given a web page
to display information about the project.

\textit{Reproducibility} is only partially considered in the 30 graded software
products. Only four products (uDig, NumPy, shapely, GDAL/OGR) provide
development setup information. In particular, shapely recommends the usage of a
virtual development environment. GDAL/OGR includes a Vagrantfile, which enables
the user to have access to a functioning virtual machine, loaded with the
project source and tools as configured by Vagrant \citep{vagrant}.

Access to sample data is provided by 24/30 projects. This sample data can be
used in the getting started tutorial, or simply to illustrate the format of the
data and to provide sample data for the user to play with. Sample data along
with a getting started tutorial (see usability) adds to reproducibility (and
correctness) since the output can be checked against what is stated in the
tutorial. However, sample data is often not comprehensive with respect to the
product's functionality, so one cannot fully grade correctness using sample
data. To fully grade correctness, a product must use a comprehensive test suite,
as discussed in the results summary for correctness.

\begin{figure}[ht]
\centering
\includegraphics[width=0.7\textwidth]{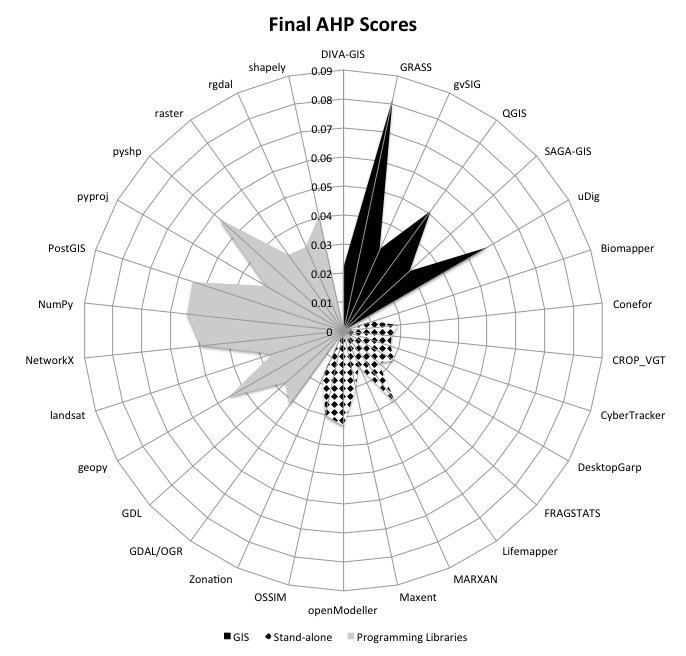}
\caption{Final AHP results.}
\label{fig:ahp-final}
\end{figure}

Once the grading has been finished, the overall impression of the products
performance on all software qualities is evaluated using AHP with equal weights
between all qualities, as shown in Figure~\ref{fig:ahp-final}. Stand-alone
tools' AHP grades are lower relative to the other two sets. This ranking is due
to the generally poor performance of stand-alone tools on installability,
correctness, maintainability, reusability, portability, understandability,
transparency and reproducibility. Part of the reason for the relatively poor
performance may be that these products have fewer developers.

\section{Recommendations}\label{recommendations}

Our recommendations assume the ideal case where the developers have the desire,
time and resources to aim for high quality.  That is, in the terminology of
\citet{gewaltig-neuroscience}, the software is intended to be user ready, as
opposed to review ready, or research ready.  Not all developers will require a
high grade on the template in~\ref{full-template}.  However, if the work will be
used for decisions that impact health, safety or financial concerns, or if the
project is to be maintained going forward, then high quality should be the goal.
Moreover, if the results obtained with the software are to be reproducible,
documentation has a critical role.

An example from GIS that stands out on all measures is GRASS.  Developers on
other projects should look to GRASS as an example to emulate.  This advice
applies to projects outside of the GIS domain.  In a paper studying SC software
in the domain of oceanography, using the same methods as used in this paper, one
of the recommendations was for oceanography developers to follow the example set
by GRASS \citep{SmithEtAl2015-OS-TR}.  The success of GRASS is, of course, based
on the hard work of the dedicated individuals that have contributed to it.
However, the success should also be attributed to the existence of a clear
software development process and an infrastructure of development support tools.

The full grading template in the Appendix should be taken as a set of criteria
for developers of SC software to consider to ensure best development practices
and product quality.  Considering all of the items on the list is recommended,
but based on the above results for the GIS domain, the authors have three main
recommendations for developers:

\begin{enumerate}

\item \textbf{Ensure the project has a requirements specification document.}
  Correctness is a quality on which many software products suffered in our
  study. By definition, correctness requires a specification. For developers to
  claim correctness, they must have complete, consistent, unambiguous, and
  verifiable specifications detailing the operation of the software product. In
  this instance, if the software was graded more leniently, perhaps some of the
  more extensive and complete user manuals could be seen as requirements
  specification documents.  Sometimes, they even included mathematical
  background. However, the nature of user manuals is to teach the end user about
  how to use the software product, not to provide requirements. The move to
  incorporate requirements is facilitated by the progress on this topic in SE. A
  structured template for requirements specification for SC software is provided
  by \citet{smith-lai-req-template}. Formal specification can also be applied
  with tools such as Frama-C \citep{framac} for C or JML \citep{jml} for Java.

\item \textbf{Provide multiple support methods.} Support for the product should
  not simply consist of directly emailing the customers. Mailing lists are
  better since they can be public, have been in use for many years and are
  relatively simple. Static methods of support such as an FAQ page or .hlp file
  (obsolete, Windows help format) are also useful, but do not allow for ad hoc
  support requests by users. ``Alternative'' methods of support should make
  support requests easier, and allow any person with the knowledge to respond.
  Some ideas for addtional support methods include an IRC channel, a Stack
  Exchange (\url{http://stackexchange.com/}) tag for new questions, or opening
  the issue tracker up for support requests. Normally, an issue tracker is only
  for bug reports, but allowing support requests to be added via the issue
  tracker gives users another way to contact developers and to get support for
  an issue with the product. This adds to the usability of the product because
  simple support makes the product simpler to use. Opening the issue tracker to
  users can assist the developers with maintainability (finding bugs), usability
  (design visibility issues), and other quality improvements. Not all of the
  above measures are necessary, especially if the software product has
  relatively few users or features. In the end, the developers for each software
  package needs to determine the appropriate level of support for their project.

\item \textbf{Design product websites for maximum transparency; for open source
    projects, provide a developer's guide.} Transparency of a product is
  important for developers because users with different backgrounds and
  intentions will be looking for information. Transparency played a large part
  in how quickly we could grade each product. Developers can make essential
  information about the project visible by creating well-designed and usable
  websites. Simple HTML websites are easy to maintain, and their design is
  straightforward.  Web platforms such as Wordpress \citep{wordpress} make
  creating and administering a blog and page style product website straight
  forward. There also exists full web solutions (like GitHub or SourceForge) to
  display product information and host source code, and other assets such as
  user manuals and issue trackers. For example, we used GitHub to host our
  project results summaries during the creation of this paper
  (\url{https://github.com/adamlazz/DomainX}).  When developers start to mix two
  or three of the above methods for their own project, transparency is greatly
  reduced. Developers are tasked with keeping multiple sites up to date while
  developing the product. If the multiple sites are not up to date, the user
  might be misguided, and the management of the product suffers.

  Transparency is especially important to consider for new team members or users
  that choose to look at the source and edit or contribute new code. The
  product's lead developers should create developer's guides as reference
  materials for these new developers. Ideally, all aspects of product
  development are represented. Information on the current state of development,
  product roadmap, design, and contribution guidelines for adding new code
  should all be included. These contribution guidelines can include any explicit
  coding standards or version control processes (e.g.\ creating a new branch for
  the patch changes). These processes increase maintainability because the
  developers have created a plan to execute when maintaining the source.

\end{enumerate}

Once these steps have been taken, we would further recommend the use of a
virtual development environment to ease reproducibility.  These are quite simple
to create nowadays, and ensures that developers' and testers' environments are
fully controlled.  This makes it simple for new developers of the product to set
up their development environment.  While this recommendation mainly concerns
developers, it is also possible that this environment can be used by end users
for a complete, isolated view of the product that requires no set up from the
user.

\section{Conclusions}\label{conclusions}

To provide feedback to the GIS software community, we systematically graded 30
software products associated with the domain. Using a multicriteria decision
making method, AHP, we performed a pairwise analysis between each software
product.  The results were summarized and interpreted for trends.

For the state of practice in GIS software we found the following
positive trends among the graded software:

\begin{itemize}
\item Products rarely have installation problems or problems with initial
  testing.

\item Projects handle garbage input without problems, such as crashing the
  program or errantly proceeding with bad input. All GIS software products
  surveyed had some error handling, which adds to their robustness.
\end{itemize}

Our survey found the following negative trends:

\begin{itemize}

\item Developers rarely explain the background knowledge or fully explain the
  intended behaviour of the product with a requirements specification document.
  Without a complete specification document, the product cannot adequately be
  judged on correctness.

\item Ideal or expected user characteristics are rarely stated, which makes it
  difficult for the user to determine if the product is right for them.

\item Instructions for validating or checking installation to ensure it works
  correctly are rarely included in the graded software. If the user is
  unfamiliar with the software product, this information would be helpful to
  them.

\item For people that want to contribute to the source code, identification of a
  coding standard should be provided, and there should be comments in the code
  indicating ``what'' is being done, but not ``how'' it is being done.  Proper
  code documentation should include pointers to more information on the
  algorithms used in the code to improve the understandability of the software.

\item Evidence that performance or maintainability are considered is rare. Lack
  of this information hurts the user's impression of the product for these
  qualities.

\item Though not a part of the software itself, the supporting web sites are
  still a part of the product. Having multiple web sites serving separate
  functions hurts transparency of the project. For example, having separate
  sites for the main product page, a repository site and a wiki site means users
  must hunt for information online that would be better gathered from a single
  well-designed web site.

\end{itemize}

\section{Acknowledgments}\label{acknowledgments}

The authors acknowledge the time and effort of fellow team members Vasudha
Kapil, Sun Yue and Zheng Zeng for their assistance in the project. In
particular, Sun Yue for development and initial documentation for a Java program
automating AHP pairwise comparisons from software grading scores.


\bibliographystyle{elsarticle-harv}
\bibliography{references}{}





\newpage

\appendix


\section{Full Grading Template}\label{full-template}

The table below lists the full set of measures that are assessed for each
software product.  The measures are grouped under headings for each quality, and
one for summary information.  Following each measure, the type for a valid
result is given in brackets.  Many of the types are given as enumerated sets.
For instance, the response on many of the questions is one of ``yes,'' ``no,''
or ``unclear.''  The type ``number'' means natural number, a positive integer.  The
types for date and url are not explicitly defined, but they are what one would
expect from their names.  In some cases the response for a given question is not
necessarily limited to one answer, such as the question on what platforms are
supported by the software product.  Case like this are indicated by ``set of''
preceding the type of an individual answer.  The type in these cases are then
the power set of the individual response type.  In some cases a superscript $^*$
is used to indicate that a response of this type should be accompanied by
explanatory text.  For instance, if problems were caused by uninstall, the
reviewer should note what problems were caused.  An (I) precedes the question
description when its measurement requires a successful installation.

\begin{longtable}{p{0.95\textwidth}}
  \caption{Grading Template}   \label{table:TemplateFull}\\
  \toprule
  \textbf{Summary Information}\\
  \midrule
  Software name? (string)\\
  URL? (url)\\
  Educational institution (string)\\
  Software purpose (string)\\
  Number of developers (number)\\
  How is the project funded (string)\\
  Number of downloads for current version (number)\\
  Release date (date)\\
  Last updated (date)\\
  Status (\{alive, dead, unclear\})\\
  License (\{GNU GPL, BSD, MIT, terms of use, trial, none, unclear\})\\
  Platforms (set of \{Windows, Linux, OS X, Android, Other OS\})\\
  Category (\{concept, public, private\})\\
  Development model (\{open source, freeware, commercial\})\\
  Publications using the software (set of url)\\
  Publications about the software (set of url) \\
  Is source code available? (\{yes, no\})\\
  Programming language(s) (set of \{FORTRAN, Matlab, C, \CC, Java, R, Ruby,
  Python, Cython, BASIC, Pascal, IDL, unclear\})\\

  \midrule
  \textbf{Installability} (Measured via installation on a virtual machine.)\\
  \midrule

  Are there installation instructions? (\{yes, no\})\\
  Are the installation instructions linear? (\{yes, no, n/a\})\\
  Is there something in place to automate the installation? (\{yes$^*$, no\})\\
  Is there a specified way to validate the installation, such as a test suite? (\{yes$^*$, no\})\\
  How many steps were involved in the installation? (number)\\
  How many software packages need to be installed before or during installation?
  (number)\\
  (I) Run uninstall, if available. Were any obvious problems caused? (\{unavail, yes$^*$, no\})\\
  Overall impression? (\{1 .. 10\})\\

  \midrule
  \textbf{Correctness and Verifiability}\\
  \midrule

  Are external libraries used? (\{yes$^*$, no, unclear\})\\
  Does the community have confidence in this library? (\{yes, no, unclear\})\\
  Any reference to the requirements specifications of the program?
  (\{yes$^*$, no, unclear\})\\
  What tools or techniques are used to build confidence of correctness? (string)\\
  (I) If there is a getting started tutorial, is the output as expected? (\{yes, no$^*$, n/a\})\\
  Overall impression? (\{1 .. 10\})\\

  \midrule
  \textbf{Surface Reliability}\\
  \midrule

  Did the software ``break'' during installation? (\{yes$^*$, no\})\\
  (I) Did the software ``break'' during the initial tutorial testing? (\{yes$^*$, no, n/a\})\\
  Overall impression? (\{1 .. 10\})\\

  \midrule
  \textbf{Surface Robustness}\\
  \midrule

  (I) Does the software handle garbage input reasonably? (\{yes, no$^*$\})\\
  (I) For any plain text input files, if all new lines are replaced with new lines
  and carriage returns, will the software handle this gracefully? (\{yes,
  no$^*$, n/a\})\\
  Overall impression? (\{1 .. 10\})\\

  \midrule
  \textbf{Surface Performance}\\
  \midrule

  Is there evidence that performance was considered? (\{yes$^*$, no\})\\
  Overall impression? (\{1 .. 10\})\\

  \midrule
  \textbf{Surface Usability}\\
  \midrule

  Is there a getting started tutorial? (\{yes, no\})\\
  Is there a standard example that is explained? (\{yes, no\})\\
  Is there a user manual? (\{yes, no\})\\
  (I) Does the application have the usual ``look and feel'' for the platform it is
  on? (\{yes, no$^*$\})\\
  (I) Are there any features that show a lack of visibility? (\{yes, no$^*$\})\\
  Are expected user characteristics documented? (\{yes, no\})\\
  What is the user support model? (string)\\
  Overall impression? (\{1 .. 10\})\\

  \midrule
  \textbf{Maintainability}\\
  \midrule

  Is there a history of multiple versions of the software?  (\{yes, no, unclear\})\\
  Is there any information on how code is reviewed, or how to contribute?
  (\{yes$^*$, no\})\\
  Is there a changelog?  (\{yes, no\})\\
  What is the maintenance type? (set of \{corrective, adaptive, perfective, unclear\})\\
  What issue tracking tool is employed? (set of \{Trac, JIRA, Redmine, e-mail,
  discussion board, sourceforge, google code, git, none, unclear\})\\
  Are the majority of identified bugs fixed? (\{yes, no$^*$, unclear\})\\
  Which version control system is in use? (\{svn, cvs, git, github, unclear\})\\
  Is there evidence that maintainability was considered in the design?
  (\{yes$^*$, no\})\\
  Are there code clones? (\{yes$^*$, no, unclear\})\\
  Overall impression? (\{1 .. 10\})\\

  \midrule
  \textbf{Reusability}\\
  \midrule

  Are any portions of the software used by another package? (\{yes$^*$, no\})\\
  Is there evidence that reusability was considered in the design? (API
  documented, web service, command line tools, ...) (\{yes$^*$, no, unclear\})\\
  Overall impression? (\{1 .. 10\})\\

  \midrule
  \textbf{Portability}\\
  \midrule

  What platforms is the software advertised to work on?
  (set of \{Windows, Linux, OS X, Android, Other OS\})\\
  Are special steps taken in the source code to handle portability? (\{yes$^*$,
  no, n/a\}) \\
  Is portability explicitly identified as NOT being important? (\{yes, no\})\\
  Convincing evidence that portability has been achieved? (\{yes$^*$, no\})\\
  Overall impression? (\{1 .. 10\})\\

  \midrule
  \textbf{Surface Understandability} (Based on 10 random source files)\\
  \midrule

  Consistent indentation and formatting style? (\{yes, no, n/a\})\\
  Explicit identification of a coding standard? (\{yes$^*$, no, n/a\})\\
  Are the code identifiers consistent, distinctive, and
  meaningful? (\{yes, no$^*$, n/a\})\\
  Are constants (other than 0 and 1) hard coded into the program? (\{yes$^*$, no, n/a\})\\
  Comments are clear, indicate what is being done, not how? (\{yes, no$^*$, n/a\})\\
  Is the name/URL of any algorithms used mentioned?
  (\{yes, no$^*$, n/a\})\\
  Parameters are in the same order for all functions? (\{yes, no$^*$, n/a\})\\
  Is code modularized? (\{yes, no$^*$, n/a\})\\
  Descriptive names of source code files? (\{yes, no$^*$, n/a\})\\
  Is a design document provided? (\{yes$^*$, no, n/a\})\\
  Overall impression? (\{1 .. 10\})\\

  \midrule
  \textbf{Interoperability}\\
  \midrule

  Does the software interoperate with external systems? (\{yes$^*$, no\})\\
  Is there a workflow that uses other softwares? (\{yes$^*$, no\})\\
  If there are external interactions, is the API clearly defined? (\{yes$^*$, no, n/a\})\\
  Overall impression? (\{1 .. 10\})\\

  \midrule
  \textbf{Visibility/Transparency}\\
  \midrule

  Is the development process defined? If yes, what process is used. (\{yes$^*$, no, n/a\})\\
  Ease of external examination relative to other products
  considered?  (\{1 .. 10\})\\
  Overall impression? (\{1 .. 10\})\\

  \midrule
  \textbf{Reproducibility}\\
  \midrule

  Is there a record of the environment used for their development and testing?
  (\{yes$^*$, no\})\\
  Is test data available for verification?  (\{yes, no\})\\
  Are automated tools used to capture experimental context?  (\{yes$^*$, no\})\\
  Overall impression? (\{1 .. 10\})\\

  \bottomrule
\end{longtable}

\section{Summary of Grading Results}

The full gradings of the 30 GIS software products are below. The most recent
gradings are available at: \url{https://data.mendeley.com/datasets/6kprpvv7r7/1}. The column
headings correspond with the above questions from the grading template.

\begin{longtable}{p{2.5cm}p{1cm}p{1cm}p{1.5cm}p{1cm}p{1cm}p{1cm}p{4cm}}
  \toprule
  Name          & Ins & Lin & Auto & Val & Steps & Pkgs & Uninstall \\
  \midrule
  DIVA-GIS      & Yes & Yes & Yes & No & 1 & 0 & No uninstall available\\
  GRASS         & Yes & Yes & Yes & Yes & 1 & 2 & No problems\\
  gvSIG         & Yes & Yes & Yes & No & 1 & 0 & No problems\\
  QGIS          & Yes & No & Yes & No & 2 & 1 & No uninstall available\\
  SAGA-GIS      & Yes & Yes & Yes & No & 1 & 1 & No uninstall available\\
  uDig          & Yes & Yes & Yes & No & 1 & 0 & No problems\\
  Biomapper     & No & N/A & Yes & No & 2 & 0 & No uninstall available\\
  Conefor       & No & N/A & N/A & No & 2 & 0 & No uninstall available\\
  CROP\_VGT     & Yes & Yes & N/A & No & 2 & 0 & No problems\\
  CyberTracker  & Yes & Yes & Yes & No & 1 & 0 & No uninstall available\\
  DesktopGarp   & Yes & Yes & Yes & No & 2 & 0 & No uninstall available\\
  FRAGSTATS     & Yes & Yes & Yes & No & 1 & 0 & No problems\\
  Lifemapper    & No & N/A & Yes & No & 2 & 0 & No uninstall available\\
  MARXAN        & No & N/A & Yes & No & 2 & 0 & No uninstall available\\
  Maxent        & Yes & No & Yes & No & 1 & 0 & No problems\\
  openModeller  & Yes & Yes & Yes & No & 1 & 0 & No problems\\
  OSSIM         & No & Yes & Yes & No & 1 & 12 & No problems\\
  Zonation      & No & N/A & N/A & No & 1 & 0 & No problems\\
  GDAL/OGR      & Yes & No & Yes & No & 1 & 2 & No uninstall available\\
  GDL           & Yes & Yes & Yes & No & 4 & 0 & No uninstall available\\
  geopy         & Yes & Yes & Yes & No & 1 & 0 & No problems\\
  landsat       & No & N/A & Yes & No & 1 & 2 & No problems\\
  NetworkX      & Yes & Yes & Yes & No & 1 & 0 & No problems\\
  NumPy         & Yes & Yes & Yes & Yes & 3 & 0 & No uninstall available\\
  PostGIS       & Yes & No & Yes & Yes & 1 & 1 & No problems\\
  pyproj        & Yes & Yes & Yes & No & 2 & 0 & No uninstall available\\
  pyshp         & Yes & Yes & Yes & No & 1 & 0 & No problems\\
  raster        & No & N/A & Yes & No & 1 & 1 & No problems\\
  rgdal         & No & N/A & Yes & No & 1 & 1 & No problems\\
  shapely       & Yes & No & Yes & No & 1 & 1 & No problems\\
  \bottomrule \addlinespace[0.5em]
\caption{Installability grading results} \label{table:installability}
\end{longtable}
\begin{longtable}{p{2.5cm}p{2cm}p{1.5cm}p{5cm}p{2cm}}
\toprule
Name          & Std Lib & Req Spec Doc & Evidence & Std Ex\\
\midrule
DIVA-GIS      & No & No & None & Yes\\
GRASS         & Yes & Yes & Programmers guide & Yes\\
gvSIG         & No & No & No & Yes\\
QGIS          & Yes & No & Developers section & N/A\\
SAGA-GIS      & Yes & No & Doxygen & N/A\\
uDig          & Yes & No & Developers guide & Yes\\
Biomapper     & No & No & None & N/A\\
Conefor       & No & No & No & N/A\\
CROP\_VGT     & No & No & None & N/A\\
CyberTracker  & Yes & No & No & Yes\\
DesktopGarp   & Yes & No & No & Yes\\
FRAGSTATS     & Yes & No & None & Yes\\
Lifemapper    & Yes & No & pydoc & Yes\\
MARXAN        & No & No & None & N/A\\
Maxent        & No & No & None & Yes\\
openModeller  & Yes & No & None & Yes\\
OSSIM         & No & No & Doxygen & Yes\\
Zonation      & No & No & None & Yes\\
GDAL/OGR      & Yes & No & Doxygen & N/A\\
GDL           & Yes & Yes & Doxygen & N/A\\
geopy         & No & No & None & No\\
landsat       & Yes & No & Extensive documentation & No\\
NetworkX      & Yes & No & None & Yes\\
NumPy         & Yes & No & None & Yes\\
PostGIS       & Yes & No & Doxygen & Yes\\
pyproj        & No & No & Wrapper to PROJ.4 library & N/A\\
pyshp         & No & Yes & No & Yes\\
raster        & Yes & No & None & Yes\\
rgdal         & Yes & No & None & N/A\\
shapely       & Yes & No & None & Yes\\
\bottomrule
\addlinespace[0.5em]
\caption{Correctness grading results} \label{table:correctness}
\end{longtable}
\begin{longtable}{p{2.5cm}p{6cm}p{2.5cm}}
\toprule
Name            & Break during install & Break during initial test\\
\midrule
DIVA-GIS        & No & No\\
GRASS           & No & No\\
gvSIG           & No & No\\
QGIS            & No & No \\
SAGA-GIS        & No & No\\
uDig            & No & No\\
Biomapper       & No & N/A\\
Conefor         & No & No\\
CROP\_VGT       & No & No\\
CyberTracker    & No & No\\
DesktopGarp     & No & No\\
FRAGSTATS       & No & No\\
Lifemapper      & Yes, install command not given & No\\
MARXAN          & No & No\\
Maxent          & No & No\\
openModeller    & No & No\\
OSSIM           & Yes, installed wrong package & No\\
Zonation        & No & No\\
GDAL/OGR        & No & N/A\\
GDL             & No & N/A\\
geopy           & No & Yes, segfault\\ 
landsat         & No & No\\
NetworkX        & No & No\\
NumPy           & No & No\\
PostGIS         & No & No\\
pyproj          & No & No\\
pyshp           & No & No\\
raster          & No & No\\
rgdal           & No & N/A\\
shapely         & No & No\\
\bottomrule
\addlinespace[0.5em]
\caption{Reliability grading results} \label{table:reliability}
\end{longtable}
\begin{longtable}{p{2.5cm}p{4cm}p{4cm}}
\toprule
Name          & Handle garbage input & Handle line ending change\\
\midrule
DIVA-GIS      & Yes & N/A\\
GRASS         & Yes & N/A\\
gvSIG         & Yes & N/A\\
QGIS          & Yes & N/A\\
SAGA-GIS      & Yes & N/A\\
uDig          & Yes & N/A\\
Biomapper     & Yes & N/A\\
Conefor       & Yes & N/A\\
CROP\_VGT     & Yes & N/A\\
CyberTracker  & Yes & N/A\\
DesktopGarp   & Yes & N/A\\
FRAGSTATS     & Yes & N/A\\
Lifemapper    & Yes & N/A\\
MARXAN        & Yes & N/A\\
Maxent        & Yes & N/A\\
openModeller  & Yes & N/A\\
OSSIM         & Yes & N/A\\
Zonation      & Yes & N/A\\
GDAL/OGR      & Yes & Yes (in scripts)\\
GDL           & Yes & Yes (in scripts)\\
geopy         & Yes & Yes (in scripts)\\
landsat       & Yes & Yes (in scripts)\\
NetworkX      & Yes & Yes (in scripts)\\
NumPy         & Yes & Yes (in scripts)\\
PostGIS       & Yes & N/A\\
pyproj        & Yes & Yes (in scripts)\\
pyshp         & Yes & Yes (in scripts)\\
raster        & Yes & Yes (in scripts)\\
rgdal         & Yes & Yes (in scripts)\\
shapely       & Yes & Yes (in scripts)\\
\bottomrule
\addlinespace[0.5em]
\caption{Robustness grading results} \label{table:robustness}
\end{longtable}
\begin{longtable}{p{2.5cm}p{9cm}}
\toprule
Name          & Evidence of performance considerations\\
\midrule
DIVA-GIS      & No\\
GRASS         & Yes, performance-specific documentation\\
gvSIG         & No\\
QGIS          & Yes, notes in wiki on performance\\
SAGA-GIS      & No\\
uDig          & No\\
Biomapper     & No\\
Conefor       & No\\
CROP\_VGT     & No\\
CyberTracker  & No\\
DesktopGarp   & No\\
FRAGSTATS     & No\\
Lifemapper    & No\\
MARXAN        & No\\
Maxent        & No\\
openModeller  & Yes, notes in wiki on performance\\
OSSIM         & No\\
Zonation      & No\\
GDAL/OGR      & No\\
GDL           & No\\
geopy         & No\\
landsat       & No\\
NetworkX      & No\\
NumPy         & No\\
PostGIS       & Yes, performance tips in documentation\\
pyproj        & No\\
pyshp         & No\\
raster        & No\\
rgdal         & No\\
shapely       & No\\
\bottomrule
\addlinespace[0.5em]
\caption{Performance grading results} \label{table:performance}
\end{longtable}
\begin{longtable}{p{2.5cm}p{1cm}p{0.75cm}p{1cm}p{1cm}p{1cm}p{0.75cm}p{5.5cm}}
\toprule
Name          & GS tutorial & Std Ex & User Man & Look and feel & Visib Prob? & User char & Support \\
\midrule
DIVA-GIS      & Yes & Yes & Yes & Yes & No & No & Mailing list, email\\
GRASS         & Yes & Yes & Yes & Yes & No & No & Mailing Lists, forum\\
gvSIG         & Yes & Yes & Yes & Yes & No & No & Bug tracker, mailing list\\
QGIS          & No & No & Yes & Yes & No & No & Mailing Lists, Forum, StackExchange, chat\\
SAGA-GIS      & No & No & Yes & Yes & No & No & Mailing list, forum\\
uDig          & Yes & Yes & Yes & Yes & No & No & Mailing list, Issue tracker, IRC\\
Biomapper     & No & No & No & No & Yes & No & Discussion list, wiki\\
Conefor       & No & No & Yes & Yes & No & Yes & Email list\\
CROP\_VGT     & No & No & No & Yes & No & No & .hlp file, email\\
CyberTracker  & Yes & Yes & Yes & Yes & No & No & Facebook/Yahoo group, email\\
DesktopGarp   & Yes & Yes & Yes & Yes & No & No & Discussion list\\
FRAGSTATS     & Yes & Yes & Yes & Yes & No & No & FAQ, email\\
Lifemapper    & No & No & Yes & Yes & No & No & None\\
MARXAN        & No & No & Yes & Yes & No & No & Mailing List, email\\
Maxent        & Yes & No & Yes & Yes & Yes & Yes & Discussion group\\
openModeller  & No & No & Yes & Yes & No & No & IRC, email\\
OSSIM         & Yes & Yes & Yes & No & Yes$^*$ & No & IRC, Mailing list, Issue tracker\\
Zonation      & No & Yes & Yes & Yes & No & No & Issue tracker, forums, wiki, \\
GDAL/OGR      & No & No & Yes & Yes & No & No & Mailing list\\
GDL           & No & No & Yes & Yes & No & No & Docs, readme, forums\\
geopy         & Yes & Yes & Yes & Yes & No & No & Github Issues\\
landsat       & No & No & Yes & Yes & No & No & Email\\
NetworkX      & Yes & Yes & Yes & Yes & No & No & Issue tracker, mailing list\\
NumPy         & Yes & Yes & Yes & Yes & No & No & GitHub, Mailing List\\
PostGIS       & Yes & Yes & Yes & Yes & No & No & IRC, Mailing list, ticket tracker, commercial support, Stack Exchange\\
pyproj        & No & No & Yes & Yes & No & No & Issue tracker\\
pyshp         & Yes & Yes & Yes & Yes & No & No & Issue tracker, email, commercial support\\
raster        & Yes & Yes & Yes & Yes & Yes & No & None. But you can find the developers email\\
rgdal         & No & No & Yes & Yes & No & No & None explicit, email?\\
shapely       & No & Yes & Yes & Yes & No & No & Github\\
\bottomrule
\addlinespace[0.5em]
\caption{Usability grading results, $^*$ has a visibility problem with the
  settings screen layout} \label{table:usability}
\end{longtable}

\begin{longtable}{p{2.5cm}p{1cm}p{1cm}p{1cm}p{1cm}p{2cm}p{2cm}p{1cm}p{1cm}}
\toprule
Name & Mul Ver & Code rvw & Chlog & Type & Issue track tool & Bugs fixes & CVS & Evid\\
\midrule
DIVA-GIS      & Yes & N/A & Yes & N/A & Email & N/A & N/A & No\\
GRASS         & Yes & PM & Yes & C & Trac & Yes & SVN & PM\\
gvSIG         & Yes & DG & No & N/A$^*$ & N/A$^*$ & N/A$^*$ & SVN & No\\
QGIS          & Yes & Yes & Yes & C & Redmine & Yes & Git & No\\
SAGA-GIS      & Yes & Yes & Yes & C & Trac & No & SVN & Yes\\
uDig          & Yes & Yes & Yes & C & JIRA & No & Git & Yes\\
Biomapper     & Yes & N/A & Yes & N/A & Email & N/A & N/A & No\\
Conefor       & Yes & No & No & N/A & N/A & N/A & N/A & N/A\\
CROP\_VGT     & Yes & N/A & Yes & N/A & N/A & N/A & N/A & N/A\\
CyberTracker  & Yes & N/A & Yes & N/A & N/A & N/A & N/A & No\\
DesktopGarp   & Yes & N/A & No & N/A & N/A & N/A & N/A & N/A\\
FRAGSTATS     & Yes & N/A & Yes & N/A & Email & N/A & N/A & No\\
Lifemapper    & Yes & No & No & N/A & N/A & N/A & N/A & Yes\\
MARXAN        & No & N/A & No & N/A & Email & N/A & N/A & No\\
Maxent        & Yes & N/A & NC & N/A & N/A & N/A & Git & No\\
openModeller  & Yes & No & Yes & C & Sourceforge & Yes & SVN & No\\
OSSIM         & Yes & No & Yes & C & Trac & Yes & Git & Doxygen\\
Zonation      & Yes & N/A & Yes & C & Redmine & Yes & N/A & No\\
GDAL/OGR      & Yes & No & Yes & C & Trac & Yes & SVN & No\\
GDL           & Yes & Yes & Yes & C & Sourceforge & Yes & CVS & No\\
geopy         & Yes & No & Yes & A & GitHub & Yes & Git & No\\
landsat       & Yes & No & No & N/A & N/A & N/A & N/A & No\\
NetworkX & Yes & DG & Yes & C, P & GitHub & Yes & Git & No\\
NumPy         & Yes & No & Yes & C & GitHub & Yes & Git & No\\
PostGIS       & Yes & DG & Yes & C & Trac & Yes & SVN & No\\
pyproj        & Yes & No & No & C & Google Code & No & Git & No\\
pyshp         & Yes & No & Yes & C & Google Code & Yes & Git & No\\
raster        & Yes & N/A & Yes & N/A & N/A & N/A & N/A & No\\
rgdal         & Yes & No & Yes & N/A & N/A & N/A & SVN & No\\
shapely       & Yes & No & Yes & C & GitHub issues & Yes & Git & No\\
\bottomrule
\addlinespace[0.5em]
\caption{Maintainability grading results}
\label{table:maintainability}
\end{longtable}
\noindent PM means Programmer's manual, DG means Developer's Guide, NC means not complete,
C means Corrective, P means Perfective, A means Adaptive, $^*$ Need account to
view issue tracker, no software showed code clones.

\begin{longtable}{p{2.5cm}p{4cm}p{4cm}}
\toprule
Name          & Portions reused & Evid \\
\midrule
DIVA-GIS      & No & No\\
GRASS         & Yes add-ons & API documentation\\
gvSIG         & Yes extensions & No\\
QGIS          & Yes plugins & Yes plugins\\
SAGA-GIS      & Yes API & API documentation\\
uDig          & Yes plugins & Plugin documentation\\
Biomapper     & No & No\\
Conefor       & No & No\\
CROP\_VGT     & No & No\\
CyberTracker  & No & No\\
DesktopGarp   & No & No\\
FRAGSTATS     & No & No\\
Lifemapper    & Not shown & Web service\\
MARXAN        & No & No\\
Maxent        & Yes API & No\\
openModeller  & Yes this is a framework & Yes\\
OSSIM         & Yes API & No\\
Zonation      & No & No\\
GDAL/OGR      & Yes API & API documentation\\
GDL           & Yes & No\\
geopy         & No & No\\
landsat       & No & No\\
NetworkX      & Yes & No\\
NumPy         & Yes & No\\
PostGIS       & Yes & No\\
pyproj        & Yes & No\\
pyshp         & Yes & No\\
raster        & Yes & No\\
rgdal         & Yes & API documentation\\
shapely       & Yes & API documentation\\
\bottomrule
\addlinespace[0.5em]
\caption{Reusability grading results, Evid means Evidence} \label{table:reusability}
\end{longtable}
\begin{longtable}{p{2.5cm}p{3cm}p{5.5cm}p{3cm}p{1cm}}
\toprule
Name          & Platform & Port in code & Not important? & Evid \\
\midrule
DIVA-GIS      & WIN OSX & N/A & No & N/A\\
GRASS         & WIN LIN OSX & Tools to create installer/ packages & N/A & N/A\\
gvSIG         & WIN LIN & Makefile & No & No\\
QGIS          & WIN LIN OSX ANDROID & Cross platform code & N/A & No\\
SAGA-GIS      & WIN LIN & Cross platform code & Yes, with OS X & No\\
uDig          & WIN LIN OSX & Eclipse & N/A & N/A\\
Biomapper     & WIN & N/A & No & N/A\\
Conefor       & WIN LIN OSX R & Unclear & No & No\\
CROP\_VGT     & WIN & N/A & N/A & N/A\\
CyberTracker  & WIN ANDROID & N/A & N/A & No\\
DesktopGarp   & WIN & N/A & Yes. No plans for Mac/Unix & N/A\\
FRAGSTATS     & WIN & N/A & No & N/A\\
Lifemapper    & WIN LIN OSX & Python & N/A & N/A\\
MARXAN        & WIN LIN OSX & N/A & No & N/A\\
Maxent        & JAVA & Java or .bat & N/A & N/A\\
openModeller  & WIN LIN OSX & Platform specific installation & N/A & N/A\\
OSSIM         & WIN LIN OSX & Compilation steps, platform specific code & No & N/A\\
Zonation      & WIN & N/A & N/A & N/A\\
GDAL/OGR      & WIN LIN OSX & Differences in makefile & N/A & N/A\\
GDL           & LIN OSX & N/A & N/A & N/A\\
geopy         & WIN LIN OSX & Python & N/A & N/A\\
landsat       & WIN LIN OSX & R & N/A & N/A\\
NetworkX      & WIN LIN OSX & Python & N/A & N/A\\
NumPy         & WIN LIN OSX & Python & N/A & N/A\\
PostGIS       & WIN LIN OSX & Differences in makefile & No & N/A\\
pyproj        & WIN LIN OSX & Python & N/A & N/A\\
pyshp         & WIN LIN OSX & Python & N/A & N/A\\
raster        & WIN LIN OSX & R & N/A & N/A\\
rgdal         & WIN LIN OSX & R & N/A & N/A\\
shapely       & WIN LIN OSX & Python & N/A & N/A\\
\bottomrule
\addlinespace[0.5em]
\caption{Portability, Evid means Evidence} \label{table:portability}
\end{longtable}
\begin{longtable}{p{2.4cm}p{1cm}p{1cm}p{1cm}p{1cm}p{1cm}p{1cm}p{1cm}p{1cm}p{1cm}p{1cm}}
\toprule
Name          & Indent & Code std & Cons Id & Cnstnts & Cmnts & URL & Params & Mdlr & File names & Design doc \\
\midrule
DIVA-GIS      & N/A & N/A & N/A & N/A & N/A & N/A & N/A & N/A & N/A & N/A\\
GRASS         & Yes & No & Yes & No & Yes & Yes & Yes & Yes & Yes & Yes\\
gvSIG         & Yes & No & Yes & No & Yes & No & Yes & Yes & Yes & Yes\\
QGIS          & Yes & No & Yes & No & Yes & Yes & Yes & Yes & Yes & No\\
SAGA-GIS      & Yes & No & Yes & No & Yes & Yes & No & Yes & Yes & Yes\\
uDig          & Yes & Yes & Yes & No & Yes & No & Yes & Yes & Yes & Yes\\
Biomapper     & N/A & N/A & N/A & N/A & N/A & N/A & N/A & N/A & N/A & N/A\\
Conefor       & Yes & No & Yes & No & Yes & No & Yes & Yes & No & No\\
CROP\_VGT     & N/A & N/A & N/A & N/A & N/A & N/A & N/A & N/A & N/A & N/A\\
CyberTracker  & N/A & N/A & N/A & N/A & N/A & N/A & N/A & N/A & N/A & N/A\\
DesktopGarp   & N/A & N/A & N/A & N/A & N/A & N/A & N/A & N/A & N/A & N/A\\
FRAGSTATS     & N/A & N/A & N/A & N/A & N/A & N/A & N/A & N/A & N/A & N/A\\
Lifemapper    & No & No & Yes & No & Yes & No & Yes & Yes & No & No\\
MARXAN        & N/A & N/A & N/A & N/A & N/A & N/A & N/A & N/A & N/A & N/A\\
Maxent        & N/A & N/A & N/A & N/A & N/A & N/A & N/A & N/A & N/A & N/A\\
openModeller  & Yes & No & Yes & No & Yes & Yes & Yes & Yes & Yes & No\\
OSSIM         & Yes & Yes & Yes & No & Yes & No & Yes & Yes & Yes & No\\
Zonation      & N/A & N/A & N/A & N/A & N/A & N/A & N/A & N/A & N/A & N/A\\
GDAL/OGR      & Yes & No & Yes & No & Yes & No & Yes & Yes & Yes & No\\
GDL           & Yes & No & Yes & No & Yes & Yes & Yes & Yes & Yes & Yes\\
geopy         & Yes & No & Yes & No & Yes & Yes & Yes & Yes & Yes & No\\
landsat       & No & No & Yes & No & Yes & No & Yes & Yes & No & Yes\\
NetworkX      & Yes & No & Yes & No & Yes & No & Yes & Yes & Yes & Yes\\
NumPy         & Yes & Yes & Yes & No & Yes & No & Yes & Yes & Yes & Yes\\
PostGIS       & Yes & No & Yes & No & Yes & No & Yes & Yes & Yes & Yes\\
pyproj        & No & No & Yes & No & No & No & Yes & Yes & No & No\\
pyshp         & Yes & No & Yes & No & Yes & Yes & Yes & Yes & Yes & No\\
raster        & Yes & No & Yes & No & No & Yes & Yes & Yes & No & No\\
rgdal         & Yes & No & Yes & No & No & No & No & Yes & Yes & No\\
shapely       & Yes & No & Yes & No & Yes & No & Yes & Yes & Yes & No\\
\bottomrule
\addlinespace[0.5em]
\caption{Understandability grading results} \label{table:understandability}
\end{longtable}
\begin{longtable}{p{2.5cm}p{7cm}p{4cm}p{1cm}}
\toprule
Name          & Ext systems & Workflow & API\\
\midrule
DIVA-GIS      & ArcView & No & N/A\\
GRASS         & Many & Not explicit & Yes\\
gvSIG         & Other softwares in project & Not explicit & Yes\\
QGIS          & GDAL framework on OS X & Not explicit & Yes\\
SAGA-GIS      & wxWidgets & Not explicit & API\\
uDig          & Eclipse Rich Client Platform & Not explicit & Yes\\
Biomapper     & None & No & None\\
Conefor       & None & Yes$^*$ & N/A\\
CROP\_VGT     & None & Not explicit & No\\
CyberTracker  & Android/Windows phones & Not explicit & N/A\\
DesktopGarp   & Microsoft XML Parser & Not explicit & No\\
FRAGSTATS     & ERSI ArcGIS ArcInfo used in tutorial & ArcGIS & N/A\\
Lifemapper    & PostgreSQL PostGIS GISs Web etc & Not explicit & Yes\\
MARXAN        & None & No & N/A\\
Maxent        & None & No & Yes\\
openModeller  & GBIF specisLink WCS & Yes & Yes\\
OSSIM         & Plugins & Not explicit & Yes\\
Zonation      & None & Not explicit & N/A\\
GDAL/OGR      & libgdal Numpy & Not explicit & Yes\\
GDL           & In software requirements & Not explicit & N/A\\
geopy         & Many third party services & Not explicit & Yes\\
landsat       & sp rgdal & Not explicit & Yes\\
NetworkX      & NumPy SciPy GraphViz and more & Not explicit & Yes\\
NumPy         & SciPy stack & Not explicit & Yes\\
PostGIS       & PostgreSQL & Not explicit & Yes\\
pyproj        & Interface to Proj.4 library & No & Yes\\
pyshp         & None & Not explicit & Yes\\
raster        & sp & Not explicit & Yes\\
rgdal         & sp & Not explicit & Yes\\
shapely       & libgeos & Not explicit & Yes\\
\bottomrule
\addlinespace[0.5em]
\caption{Interoperability, $^*$Conefor input generating GIS plugins} \label{table:interoperability}
\end{longtable}
\begin{longtable}{p{2.5cm}p{4cm}p{4cm}}
\toprule
Name          & Dev process & External exam \\
\midrule
DIVA-GIS      & No   & 5\\
GRASS         & Yes, developer's guide    &10\\
gvSIG         & Yes, developer's guide   & 4\\
QGIS          & Yes, developer's guide & 8\\
SAGA-GIS      & Yes, developer's guide & 7\\
uDig          & Yes, developer's guide    &10\\
Biomapper     & No  & 3\\
Conefor       & No& 6\\
CROP\_VGT     & No  & 4\\
CyberTracker  & No   & 6\\
DesktopGarp   & No& 7\\
FRAGSTATS     & No  & 6\\
Lifemapper    & No & 6\\
MARXAN        & No & 4\\
Maxent        & No & 9\\
openModeller  & No   & 6\\
OSSIM         & No  & 6\\
Zonation      & No   & 4\\
GDAL/OGR      & No   & 4\\
GDL           & Yes, HACKING file & 8\\
geopy         & No  & 9\\
landsat       & No& 5\\
NetworkX      & Yes, developer's guide & 9\\
NumPy         & No  & 4\\
PostGIS       & Yes, developer's guide & 8\\
pyproj        & No & 6\\
pyshp         & No  & 9\\
raster        & No & 9\\
rgdal         & No & 4\\
shapely       & No & 9\\
\bottomrule
\addlinespace[0.5em]
\caption{Visibility grading results} \label{table:visibility}
\end{longtable}
\begin{longtable}{p{2.5cm}p{1cm}p{8cm}p{2.5cm}}
\toprule
Name          & Dev env & Ver test data & Tools capture exp context \\
\midrule
DIVA-GIS      & No & Sample data not for verification & None\\
GRASS         & No & Sample data and test suite & None\\
gvSIG         & No & Tests exist & None\\
QGIS          & No & Sample data available and test suite available & No\\
SAGA-GIS      & No & Tests available & None\\
uDig          & Yes & Sample data and test suite & None\\
Biomapper     & No & No & None\\
Conefor       & No & Sample data not for verification & None\\
CROP\_VGT     & No & No & None\\
CyberTracker  & No & Sample data not for verification & None\\
DesktopGarp   & No & No & None\\
FRAGSTATS     & No & Yes & None\\
Lifemapper    & No & Sample data not for verification & None\\
MARXAN        & No & No & No\\
Maxent        & No & Sample data not for verification & No\\
openModeller  & No & Sample data and test suite & None\\
OSSIM         & No & Yes and test suite & None\\
Zonation      & No & Sample data not for verification & None\\
GDAL/OGR      & No & Tests & Vagrantfile\\
GDL           & No & Test suite & None\\
geopy         & No & Test suite & No\\
landsat       & No & No & None\\
NetworkX      & No & Test suite & No\\
NumPy         & Yes & Test suite & None\\
PostGIS       & No & Yes test suite, make check & None\\
pyproj        & No & Test suite & None\\
pyshp         & No & Test suite & None\\
raster        & No & No & None\\
rgdal         & No & Tests available & None\\
shapely       & No$^*$ & Tests available & None\\
\bottomrule
\addlinespace[0.5em]
\caption{Reproducibility, $^*$Virtual environment preferred} \label{table:reproducibility}
\end{longtable}



\end{document}